# Using MongoDB for Social Networking Website

Deciphering the Pros and Cons

Sumitkumar Kanoje  
Dept. of Info Technology  
Maharashtra Inst. of Technology  
Pune, India  
sumitkanoje@gmail.com

Varsha Powar  
Dept. of Info Technology  
Maharashtra Inst. of Technology  
Pune, India  
varsha.powar@mitpune.edu.in

Debajyoti Mukhopadhyay  
Dept. of Info Technology  
Maharashtra Inst. of Technology  
Pune, India  
debajyoti.mukhopadhyay@gmail.com

*Abstract*—Social media is a biggest successful buzzword used in the recent time. Its success opened various opportunities for the developers. Developing any application requires storage of large data into databases. Many databases are available for the developers, Choosing the right one make development easier. MongoDB is a cross platform document oriented, schema-less database eschewed the traditional table based relational database structure in favor of JSON like documents. This article discusses various pros and cons encountered with the use of the MongoDB so that developers would be helped while choosing it wisely.

## I. INTRODUCTION

Traditionally, relational databases have been helpful for developers in developing applications, but due to its limitations like two dimensional table to represent the data, strict consistency of database transactions, more time to read & write with sometimes leading to the use of complex queries and availability of many other databases with advancement of technology has encouraged developers to use other NoSQL databases.

This paper attempts to use one of this new technology database viz. MongoDB for the development of social networking websites.

Here we discussed the various pros and cons encountered with the use of MongoDB so that developers would be helped with the use of it. It is partitioned into 6 parts. Section (1) presents introduction (2) presents social networking website components. Section (3) describes MongoDB database. Section (4) presents pros of using MongoDB database. Section (5) describes cons of using MongoDB with the applicability to social networking website. And section (6) concludes the paper.

## II. SOCIAL NETWORKING WEBSITE

### A. Background

Social networking service is platform which gives peoples opportunity to build social network or social relations among people across the globe, who shares their thoughts, interests, activities or real life connections. Early startups have grabbed this opportunity for being successful due to the high popularity and interest among people.

For discussing a good database for social networking website we must know the basic components of social networking website, will see these components in next part of this section.

### B. Components of Social Networking Websites

Below figure represents the social networking website. As it involves the various activities of several users, the main components of these websites can be considered as People/Profile, Post/Blog, Media/Image/video/Audio, Relations/Interactions and API's.

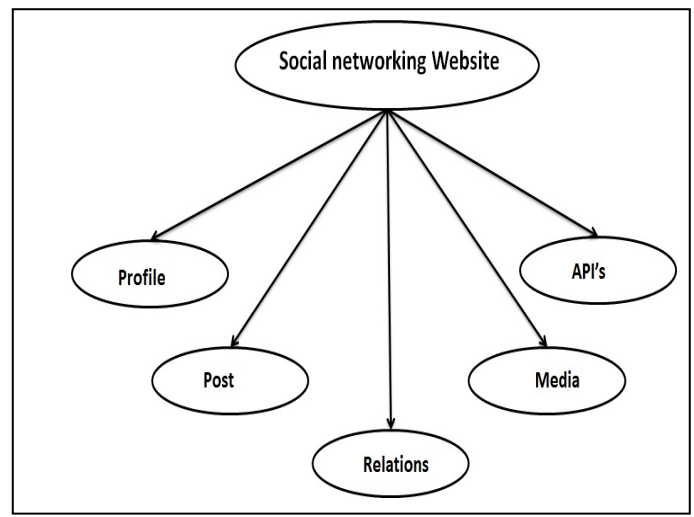

Fig. 1. Modules of Social Networking Website

*1) Profile:*

This component collects user related information viz. biographical information (name, age, sex, living), educational information and professional information. These details give identity to user on the social networking website. This identity can be used by other users to form network among users. A profile can have connections with other profiles i.e. friendship, follow or subscribe.

*2) Post:*

This component gives editing and publishing functions to users. For example user can post a blog, tweet etc. He can also publish some of his experiences.

*3) Media:*

Media includes various photos, audios and videos which a user wants to share with his connections. Media is the most important entity for a user.

*4) Relations:*

Relations or interaction can be the likes, shares and comments etc. which are received on a post or blog.

*5) API's:*

API gives social networking websites a way to interact with the other websites or apps. API cannot be ignored as these are the most important entity responsible for the success of most social networking websites.

### III. COMING TOWARDS MONGODB

MongoDB is a scalable high performance open-source NoSQL database. MongoDB uses document store instead of two dimensional table structures. It was developed by the company 10gen as component of a planned platform as a service product [1]. It was being adopted as backend software by major services companies including eBay, SourceForge, FourSquare and New York Times.

The differences between MongoDB and normal Relational DBMS can be well represented in the following figure. The figure shows how Tables in RDBMS are replaced by Collections in MongoDB, Rows are replaced by JSON documents and Joins are replaced by Embedding and linking.

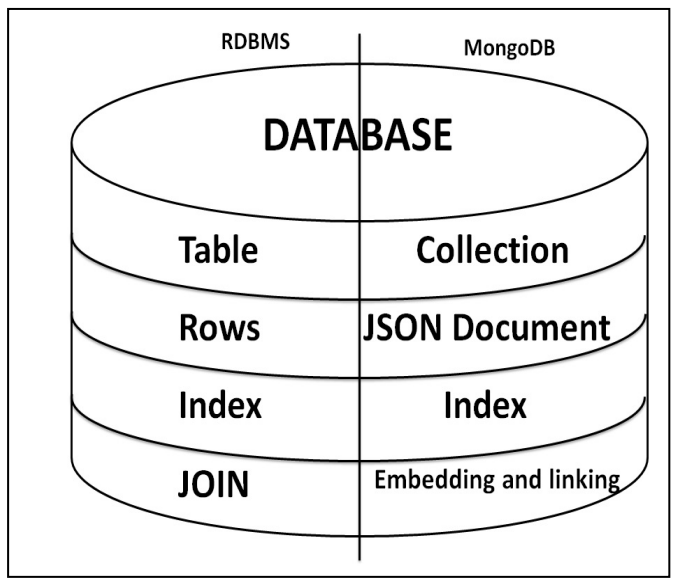

Fig. 2.  MongoDB vs. Relational Database

MongoDB is good at scaling, uses Map Reduce and has descent geospatial integration

### IV. PROS OF USING MONGODB

MongoDB has attracted many developers due to its features superior over the traditional databases. MongoDB has evolved as a new form of database which can be used by the developers. The reason why one should use MongoDB for Social Networking Websites are document oriented storage, indexes on any attribute, Replication and High Availability, Auto Sharding, Rich Queries, Fast in-place updates. These advantages are discussed in more details below.

One of the first and most promising advantages which have attracted developers toward MongoDB is the less initial development efforts needed. As MongoDB follows procedures and methods like commands for interacting with it, developers don't have to waste much of the time in learning how to write code for MongoDB database. It sounds weird but it has been observed by many developers that writing MongoDB commands is more easier than written those long SELECT * FROM like queries. Its example can be given as below [4]

```
Mongo m = new Mongo("DBServer", 27017);
    //connect database with database server and port
DB db = m.getDB("dbname");   //Open database
```

The above example shows a simple connection establishment with MongoDB database. Now to insert the data it is like

```
DBCollection coll = db.getCollection("collectionname");
BasicDBObject doc = new BasicDBObject();
Doc.put("attribute", "value");
Coll.insert(doc);
```

One more advantage of MongoDB is its good at scaling. Scaling is achieved using Sharding in MongoDB. Sharding is the process of storing data records across multiple machine nodes and is MongoDB's approach to meeting the demands of data growth. With increasing size of data, a single machine may seem insufficient to store the data. Also it provides unacceptable read and write throughput. This problem of horizontal scaling is solved using sharding process. With Sharding, we add more machines to support data growth and demands of read and write operations [3].

MongoDB is Document Oriented Database in which one collection holds different documents, number of fields. Content and size of document can differ from one document to another. Even the user can change/add/delete any of these documents at any time, this feature makes MongoDB Schema-less.

Deep query ability of MongoDB is outstanding. MongoDB supports dynamic queries on documents using a document based query language that is nearly as powerful as SQL. This allows documents retrieval from database in short duration.

Indexes support the efficient execution of queries in MongoDB. MongoDB defines indexes at the collection level and supports indexes on any field or sub-field of the documents in a MongoDB collection.

MongoDB's support for API development can be additional advantage in selecting MongoDB for developing a Social Networking Website [6]. Its document store structure suits the object structure that gets returned or processed by an API in response to GET or POST requests. Such structure is well used by Social Networking giant Facebook.

Other advantages of using MongoDB include easy replication, MapReduce, clustering etc.

## V. CONS OF USING MONGODB

Although we considered MongoDB as suitable for social networking website, there is another side of the coin that should be considered while selecting MongoDB for any social networking project. Many developers have realized this and shifted from using MongoDB to other databases within one year of their project startup [7]. So going to the final decision regarding database selection without considering these cons would lead to risks. The cons are discussed below.

Performing JOIN query is one of the crucial points which have made developer think again before using MongoDB. Every application at some point needs to perform JOIN operations. But in case of MongoDB which uses document like structure. If one need information from one table to filter the information from another table (i.e. a join), then MongoDB is going to work against us. In a SQL database we can easily get the info from multiple tables with a single JOIN query. In MongoDB we need multiple queries and join the data manually within our code (which may cause slow & ugly code, also reduced flexibility when the structure changes) to get data from multiple collections. Any speeds advantage that mongo gives us in pulling from a single collection will quickly be negated by making multiple round trips to the database [8].

MongoDB supports indexes on any field or sub-field of the document; even it supports compound and multi-key indexes. But in some of the practical implementations it has been found that setting up indexes on many fields or attributes takes up a lot of RAM space. Actually these are B-tree indexes and if we have many, we might run out of system resources really fast.

There are some concurrency issues in MongoDB, when we perform a write operation in MongoDB it creates a lock on the entire database, not just the affected entries and not just for a particular connection, thus lock blocks not only write operation but also read operations.

Operations are not automatically treated as transactions by MongoDB. We have to manually choose to create a transaction, verify it manually, and then manually commit or rollback for ensuring data integrity upon create/update operations. Operations on a single document are always atomic with MongoDB databases; however, operations which involve multiple documents are not atomic, operations are often referred to as "transactions". Many practical use cases are supported by single-document atomicity as documents can be fairly complex and contain multiple "nested" documents.

Some other issues that can be considered at lower extent are less documentation, smaller community or experience with a document store database than relational database like MySQL.

## VI. CONCLUSION

This thought paper compares the pros and cons of using MongoDB database for developing Social Networking Website. After the long discussions in the above four sections we can say that MongoDB is good at some point for developing social networking website and somewhere it lags too. But that doesn't means it should never be used, it might not the optimal solution in some situations. That doesn't mean it don't have a lot potential. Our database choice should be application specific. It has been experienced by many developers that every nonstandard technology choice reduces ability to iterate quickly. Looking at the application point of view Facebook is used by a billion peoples over the world and even Facebook doesn't use one technology, so right now MongoDB might not be having that potential to compete with relational databases but has a great future ahead.